\documentclass[twocolumn,aps,prb,showpacs,preprintnumbers,amsmath,amssymb, superscriptaddress]{revtex4}

\usepackage{graphicx}
\usepackage{dcolumn}
\usepackage{bm}

\begin{document}


\title{Electronic States in One, Two, and Three Dimensional Highly Amorphous Materials: 
A Tight Binding Treatment}


\author{D. J. Priour, Jr}
\affiliation{Department of Physics, University of Missouri, Kansas City, Missouri 64110, USA}


\date{\today}

\begin{abstract}
In a tight binding framework, we analyze the characteristics of electronic states in 
strongly disordered materials (hopping sites are placed randomly with no local order) 
with tunneling matrix elements decaying exponentially in the atomic separation with various 
decay ranges $l$ examined.  We calculate the density of states (DOS) and the Inverse Participation 
Ratio (IPR) for amorphous atomic configurations in one, two, and three dimensions.  
With a finite size scaling analysis of the IPR statistical distributions,  
it is shown that states are either extended or localized for a particular 
energy, and phase portraits for wave functions are obtained showing extended 
and localized behavior in the thermodynamic limit.
While we conclude that all states are localized in 1D, in the 
2D case there is a threshold for $l$ above which some eigenstates appear to be 
extended and below which wave functions are entirely localized.  For 3D geometries, 
there are two mobility boundaries flanking an intermediate range of 
energies where states are extended with eigenstates localized for energies 
above or below this range.  While a zone of extended states persists even for 
very short $l$, the width of the region tends to zero exponentially (i.e. scaling as $e^{-A/l}$) 
for very small decay length scales.
\end{abstract}

\maketitle


\section{Introduction and Theoretical Framework}
A periodic crystal in the absence of disorder supports extended Bloch 
waves within the bounds of energy bands~\cite{Bloch}.  However, the 
nature of the electronic states
in a strongly disordered (i.e. amorphous) material where symmetry with 
respect to discrete translations is absent, is a more 
subtle question.  Our aim is to examine the effect of very strong disorder on transport
characteristics.  

Operating in the framework of a tight binding model where 
electrons are localized in atomic sites or dopant impurities, we examine amorphous materials in one,   
two, and three dimensions where the locations of atoms are taken to be 
uncorrelated and randomly distributed within the medium. Atomic or dopant configurations 
with no local ordering of the sites are described as gas-like disorder~\cite{Ziman} 
with relevance to the transport characteristics associated with expanded alkali 
metals~\cite{Hensel} as well as 
impurity bands in silicon.  The characteristics of exciton states with respect to localization 
have been 
examined in the context of similar types of disorder~\cite{Blumen}. 
In broader generality, formal analytical and computer 
studies have calculated the Density of States in amorphous materials with no correlations 
among the site positions~\cite{Ching,Logan1,Logan2,Logan3}

Disorder, even in regular lattices, may be manifest as random site energies which 
can disrupt the extended character of itinerant states and thereby create conditions 
for localization.  Our aim is to examine strongly disordered materials and the 
properties of the associated electronic states with respect to localization.
However, we do not introduce a random site energy, and in this sense our work is 
complementary to studies where random potentials are superimposed on sites in a periodic
crystalline geometry~\cite{Brndiar,Wobst,Mildenberger,Bauer,Schreiber}.
Instead of examining a system on a regular lattice geometry, we calculate 
electronic wave functions for the fully amorphous case 
and examine how strong positional disorder,
in conjunction with tunneling matrix elements which decay exponentially 
in the separation of neighboring hopping sites, affects the 
characteristics of eigenstates with respect to localization.

Off-diagonal disorder which enters in random variations in tunneling matrix elements controlling 
hopping among sites in the 
absence of a random on-site potential, has been of interest  
since an early study by F.~J. Dyson~\cite{Dyson,Gade,Parshin,Cerovski,Takahashi,Xiong,Garcia}. 
Calculations related to the density of states~\cite{Logan1,Logan2,Logan3} and aspects of 
localization~\cite{Ching,Blumen,Logan4} have been carried out
in the context of a 3D gas-like tight binding model. 
We report on results relevant to the aim of finding out by direct calculation in a large-scale statistical study 
the extent to which wave functions are localized in amorphous materials and under what 
conditions eigenstates are extended.  
Moreover, we obtain
phase portraits showing for different energies and 
ranges $l$ of the hopping integral domains of extended and localized states.

In this work, we use for the tight binding Hamiltonian 
\begin{align}
{\mathcal H} = -\frac{1}{2} t_{0} \sum_{i=1}^{N} \sum_{j \neq i} V(r_{ij}) (\hat{c}_{i}^{\dagger} \hat{c}_{j} + 
\hat{c}_{i} \hat{c}_{j}^{\dagger}) 
\label{Eq:eq1}
\end{align}
where the sum over the index ``$i$'' ranges over the $N$ particles contained in the simulation 
volume, we take the hopping parameter $t_{0}$ to be 3.0 electron volts, and the factor of ``1/2'' 
compensates for multiple counting of the hopping terms between atoms.  The creation 
and destruction operators $\hat{c}^{\dagger}$ and $\hat{c}$ create and destroy occupied electronic orbitals at 
sites indicated by the subscript.  For the hopping integral, we use $V(r_{ij}) = e^{-\gamma r_{ij}/s}$, 
where $r_{ij}$ is the separation between sites $i$ and $j$,
$s = \rho^{-1/D}$ is the typical inter-orbital separation, $\rho$ 
is the volume density of sites, $D$ is the dimensionality of the system, and  
$\gamma$ is a dimensionless parameter.  With the hopping integrals expressed in this manner and the 
length scale for the decay of the hopping matrix element thus being $l = s/\gamma$, large/small $\gamma$ 
values correspond to decay lengths small/large in relation to the typical interatomic separation.
For the sake of convenience, we rescale coordinates such that $\rho = 1$, with a simple 
inverse relationship  $l = \gamma^{-1}$ among the hopping length scale $l$ and decay    
parameter $\gamma$ 
and $V(r_{ij}) = e^{-\gamma r_{ij}}$ for the dependence of the tunneling matrix element.
Although the hopping integral $V(r_{ij})$ is finite in range by virtue of the exponential decay,
we nevertheless take into consideration hopping among all pairs of orbitals contained in the system of 
$N$ sites with only a negligible increase (i.e. a contribution on the order of $N^{2}$) where direct 
diagonalization, which scales as $N^{3}$ and constitutes the main computational bottleneck, 
is used to calculate the electronic states.

Our Hamiltonian does not incorporate perturbations of the on-site energy term 
(for convenience all site energies are set to zero), and disorder instead 
enters in the off-diagonal tunneling matrix elements.  Thus, our program in this 
work is to investigate the effect of positional disorder itself on transport 
characteristics.  Although the treatment is nonperturbative with no local order in the 
locations of the sites in the amorphous systems we examine, the decay length scale $l$ 
of the hopping integral serves to parameterize the disorder strength.
For large $l$ (small $\gamma$), the tunneling is long-ranged and connects sites 
to many neighbors, effectively averaging over the hopping rates to and from 
many orbitals and thereby somewhat 
muting the effect of positional disorder  On the other hand, if $l$ is small 
relative to the typical inter-site spacing $\rho^{-1/D}$ (i.e. for large $\gamma$), the
tunneling is preponderantly to the very closest neighbors; moreover, even small 
fluctuations in the locations of nearest neighbors impact charge transport through the 
site to a significant degree via the exponential dependence of 
the hopping integral.  Hence, disorder is in a sense amplified as $\gamma$ is 
increased and muted when $\gamma$ is small.

In calculating the tight binding wave functions, we examine a $L^{D}$ supercell, where many  
system sizes are considered in order to perform finite size scaling and determine the 
degree to which eigenstates are localized in the bulk limit.
Energies obtained by diagonalizing the tight binding (Hermitian) Hamiltonian are used to construct 
the global density of states (DOS), while the eigenstates themselves are retained for analysis to 
characterize the electronic wave functions with respect to localization with the aid of
a single quantity known as the 
Inverse Participation Ratio (IPR), ${\displaystyle \sum_{i=1}^{N} \lvert \psi_{i} \rvert^{4} }/
\big ({\displaystyle \sum_{i=1}^{N}} \lvert \psi_{i} \rvert^{2} \big )^{2}$.
The participation ratio shows distinct behavior depending on whether the wave function 
is confined to a small volume or 
spread out over a larger region, and hence more extended in character.  While in the former case the IPR is 
finite and tends to be relatively large, the participation ratio is smaller for broader electronic states, and 
approaches zero as the wave function becomes spread over a bulk system in the case of a genuinely 
extended state.  This dichotomy for extended \textit{vis \`{a} vis} localized states makes the IPR a 
useful diagnostic parameter in the context of theoretical calculations where often 
the participation ratio provides information as to the extent of localization of carrier  
wave functions in specific locations in an amorphous geometry or in certain energy ranges of 
a band structure~\cite{Cai,Turek,Blaineau,Holender,Koslowski}.

In this work, we aim to determine for very large 1D, 2D, and 3D amorphous systems the prevalence
of extended states, and we construct phase portraits showing where 
wave functions are localized and where states are extended.
  The random character of the disorder precludes a direct observation of the 
evolution of the characteristics of electronic states as the system size $L$ is increased.  
However, there remains the possibility of determining in a statistical sense the localization 
characteristics of the amorphous medium by calculating the Inverse Participation Ratio 
histogram.  Shifts in the weight of the IPR probability distribution with increasing $L$
provide information as to how many of the electronic states ultimately are localized and what        
portion are genuinely extended.  For a specific hopping integral decay parameter 
$\gamma$ we find the status of electronic states with respect to 
localization to be determined exclusively by the energy eigenvalue, with no situation arising in which localized and 
extended states exist simultaneously for the same infinitesimal energy interval.
With a finite size scaling analysis of the Participation ratio 
histogram calculated for various system sizes, we extrapolate to the bulk limit and determine in a rigorous fashion 
energies where electronic states are localized, and eigenenergies supporting extended states.

By repeating the calculation for different $\gamma$, we obtain phase portraits 
showing regions of localized and extended wave functions.
The IPR probability distributions represent an intermediate stage in the determination of energy ranges
corresponding to localized or extended states in the bulk limit, but are of interest in themselves
and highlight salient qualitative trends.  A participation ratio histogram which does not change either 
in its shape or position with increasing system size signifies that all states encompassed in the 
distribution are localized, and the IPR distribution may be regarded as a bulk characteristic where 
$L \gg \xi$ with $\xi$ being the localization length scale of the eigenstates.  At the opposite 
extreme is a Participation ratio histogram which shifts in its entirety toward smaller IPR values  
with increasing $L$.  The steady transfer of a large share of statistical weight toward 
even smaller participation ratios implies the electronic states are preponderantly extended.  A possibility interpolating 
between the two extremes arises if a part of the IPR distribution converges as would be 
expected for a set of localized states while at the same time a portion of the histogram 
separates from the main envelope of the distribution and is conveyed toward lower IPR values.

To sample realizations of disorder from the appropriate statistical 
distribution, it is important to generate random configurations of atoms in an 
unbiased way.  For stochastic input, we use a Mersenne Twister algorithm to minimize correlations 
among successively generated random numbers and to ensure the period of the pseudorandom 
sequence far exceeds the quantity of random numbers used over the course of the simulations. 
For continuously distributed hopping sites, the number $N$ of sites in the simulation volume 
must in general vary from one sample to the next due to 
statistical fluctuations.  The integer value closest 
to the mean occupancy $N_{\mathrm{av}} = \rho L^{D}$
is a convenient initial choice, and random variations in the number of particles in the 
simulation volume are taken into account with a sequence of stochastically driven 
attempts either to raise or lower $N$. The latter are part of an importance sampling scheme similar to that used to 
derive the Metropolis Criterion~\cite{met} at the heart of Monte Carlo simulations 
which sample the Boltzmann distribution in the calculation of thermodynamic variables.

To determine the probability of $N$ sites in $v = L^{D}$, we divide $v$
into $M$ sub-volumes of equal size where $\Delta v = v/M$.  For large $M$ the likelihood 
of multiple occupancy in any of the sub-volumes is very small relative to the chance of having one or 
zero sites in a subdivision; in the small $\Delta v$ limit, the single occupancy 
probability is $\rho \Delta v$, with $(1 - \rho \Delta v)$ being the complementary likelihood of null
occupancy.  Hence, the probability the entire system is devoid of hopping sites is $(1 - \rho v/M)^{M}$, which 
becomes $e^{-\rho v}$ for $M \rightarrow \infty$.  For single occupancy, adopting a prefactor $M$ to 
take into consideration that the site may reside in any of the $M$ sub-volumes, yields $M(\rho v/M)(1 - \rho v/M)^{M-1}$, 
which becomes $\rho v e^{-\rho v}$ in the $\delta v \rightarrow 0$ limit.  Similar logic is used for 
general case, and the probability for having exactly $N$ sites in the simulation volume is
$P(N) = e^{-\rho v} (\rho v)^{N}/N!$ where $N!$ is a combinatorial factor to compensate for multiple counting.

To generate a realization of disorder, a succession of attempts (a number of moves in the vicinity of 
$N_{\mathrm{av}}$ is sufficient to achieve ergodicity) is made to raise or lower the occupancy 
number $N$, where the choice to increase or decrease $N$ is randomly determined.  For increments 
from $N$ to $N+1$, the relevant criterion is the probability ratio $r_{+} \equiv p(N+1)/p(N) = \rho v/(N+1)$, and the   
change is accepted is $X_{r} < r_{+}$, where $X_{r}$ is a random number sampled uniformly from the interval
$[0,1]$.  Similarly, decreasing $N$ to $N-1$ occurs if $X_{r} < r_{-}$ where $r_{-} = p(N-1)/p(N) = N/\rho v$.
With $N$ properly sampled, $D$ Cartesian coordinates for each site location are chosen independently 
(and at random with uniform probability density) from the 
interval $[0,L]$.  

The coordinates for each of the $N$ sites enter in the construction of the Hamiltonian matrix (in 
the context of the tight binding model given in Eq.~\ref{Eq:eq1}),  which is diagonalized for the 
eigenenergies and eigenstates; periodic  boundary conditions are implemented
to mitigate finite size effects.          
For the purpose of the DOS calculations, $5 \times 10^{5}$ energy
eigenvalues  are sampled; in obtaining the IPR statistical distributions, where eigenstates are
used to calculate participation ratios
with a concomitant increase in the computational burden, $10^{5}$
wave functions are retained. 

In each section of this work, we present and discuss results for 1D, 2D, and 3D systems; for 
each dimensionality, we examine a range of tunneling matrix element decay parameters $\gamma$
which in a sense govern the strength of the disorder.  In Section II, we examine the Density of 
States for energy eigenvalues. Inverse Participation Ratio (IPR) statistical 
distributions are discussed in Section III with IPR histograms displayed for various system sizes.
In Section IV, the channel averaged participation ratios are shown;  we argue that IPR channel averages are 
representative of a particular energy and hence may be used to determine how the IPR scales with 
system size.  Ultimately, in Section V, a finite size scaling analysis of the participation ratio 
channel averages is used to extrapolate to the thermodynamic limit and thereby yield the mean bulk IPR. 
The latter, used to construct phase portraits showing where states are extended or localized for 
different values of $E$ and $\gamma$, indicates for $D = 3$ a zone of extended states in the midst of a 
region of localized states and flanked by two mobility boundaries.  Although a finite fraction of the 
states are extended even for large $\gamma$ where disorder fluctuations are very important, the 
width of the interval of energies $w(\gamma)$ where states are extended decreases rapidly for large to moderate $\gamma$, 
with an asymptotically exponential dependence $w(\gamma) \sim e^{-A\gamma}$.
Conclusions are presented in Section VI.

\section{Energy Density of States}

\begin{figure}
\includegraphics[width=.49\textwidth]{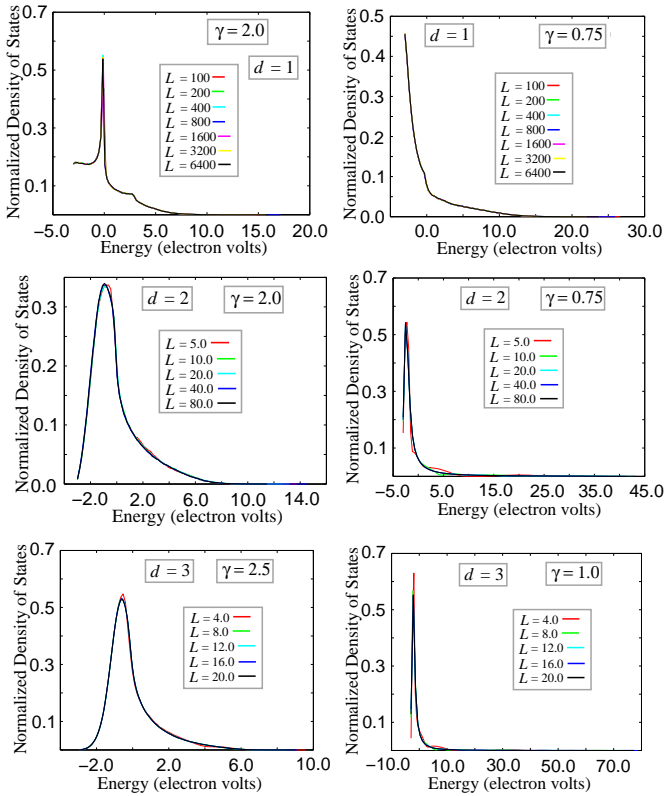}
\caption{\label{fig:Fig1} (Color Online) Density of states plotted for (top to bottom) 
$D = 1$, $D = 2$, and $D = 3$, and for large to small $\gamma$ (left to right).  DOS
curves are plotted for various systems sizes $L$.}  
\end{figure}

\begin{figure}
\includegraphics[width=.49\textwidth]{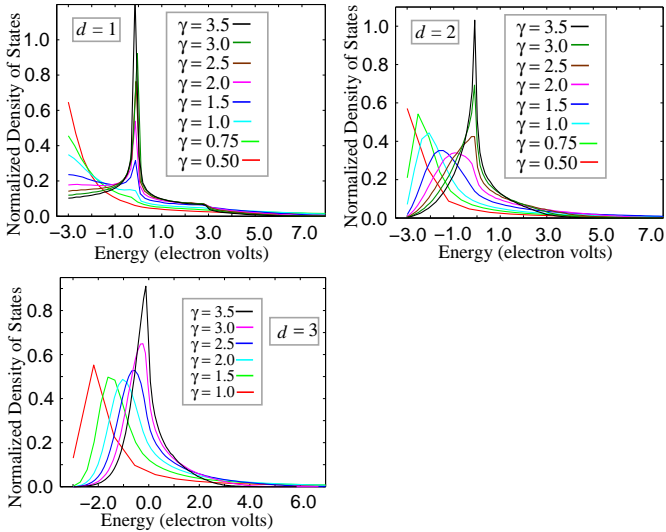}
\caption{\label{fig:Fig2} (Color Online) Density of States curves are shown together for a range of decay rates
$\gamma$, with, clockwise from left, $D = 1$, $D = 2$, and $D = 3$.}
\end{figure}

The DOS, which in principal is a continuous function $f(E)$ of 
energy for an amorphous material, may only be rendered to an approximate degree in a finite calculation.  
Accordingly, we partition the energy interval between the ground state and the uppermost 
excited state into a       
finite though reasonably large number of sub-intervals or ``bins''.   
The augmented resolution achieved with an increase in the
total number of partitions is counterbalanced with 
a rise in the magnitude of 
statistical fluctuations.  To strike a suitable balance among detail and  
noise, 500 divisions are used in preparing DOS histograms;
with a total of $5 \times 10^{5}$ eigenvalues for each DOS curve,
choosing 500 bins still allows on average 
for 1,000 data point for each partition.

We examine DOS curves for 1D, 2D, and 3D systems, for which a salient common characteristic 
is a rapid convergence of the energy 
eigenvalue statistical distributions with respect to the system 
size $L$.  In fact, as may be seen for representative cases in Fig.~\ref{fig:Fig1},
the DOS curves overlap very closely and approach the bulk limit as long as 
$L \gg \mathrm{max}[\rho^{-1/D},l]$ with the hopping range $l$ and the 
typical interparticle separation $\rho^{-1/D}$ (unity in our treatment) being relevant length scales.

Another important trend seen for all dimensionalities under consideration is a systematic 
shift of the DOS statistical weight toward lower (i.e. more negative) energies with  
decreasing $\gamma$, or increasing tunneling matrix element range $l = \gamma^{-1}$ shown. 
In Fig.~\ref{fig:Fig2} DOS curves are displayed for $D = \left \{ 1,2,3 \right \}$ for a broad 
range of $\gamma$ values with results (calculated for large $L$) 
converged with respect to the size of the system.

Although a shift of statistical weight toward negative energy values occurs 
for each dimensionality, the evolution of the DOS curves for 1D 
systems differs from that seen for $D =2$ and $D = 3$.  For  
$D = 1$, the energy probability distribution has a single maximum 
centered about $E = 0$ for $\gamma \gg 1$.  However, with increasing $l = \gamma^{-1}$,
a secondary peak appears at a negative 
energy, gaining amplitude at the expense of the probability density near the 
$E = 0$ maximum; for large enough 
$l$, the latter vanishes with only the negative energy peak remaining. 

For two and three dimensional systems, probability density is transferred from $E = 0$ to 
negative energies as in the case $D = 1$. However, instead of an intermediate transition 
to a dual peak profile, the DOS curves remain unimodal as the single maximum
migrates continuously toward lower energies, arriving at a 
negative energy for $\gamma \ll 1$.   

\section{Participation Ratio Statistical Distributions} 

\begin{figure}
\includegraphics[width=.49\textwidth]{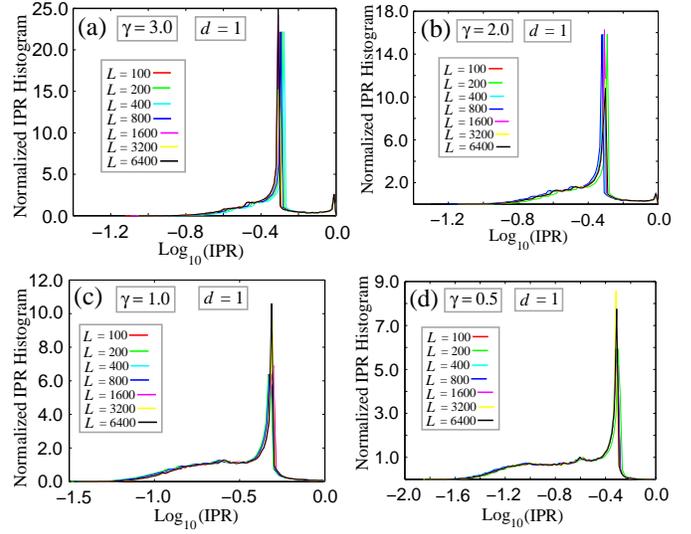}
\caption{\label{fig:Fig3} (Color Online) Inverse Participation Ratio profiles graphed for
various system sizes $L$ for hopping integral
decay rates ranging from $\gamma = 3.0$ in panel (a) to $\gamma = 0.5$ in panel (d) for 1D systems.}
\end{figure}

\begin{figure}
\includegraphics[width=.49\textwidth]{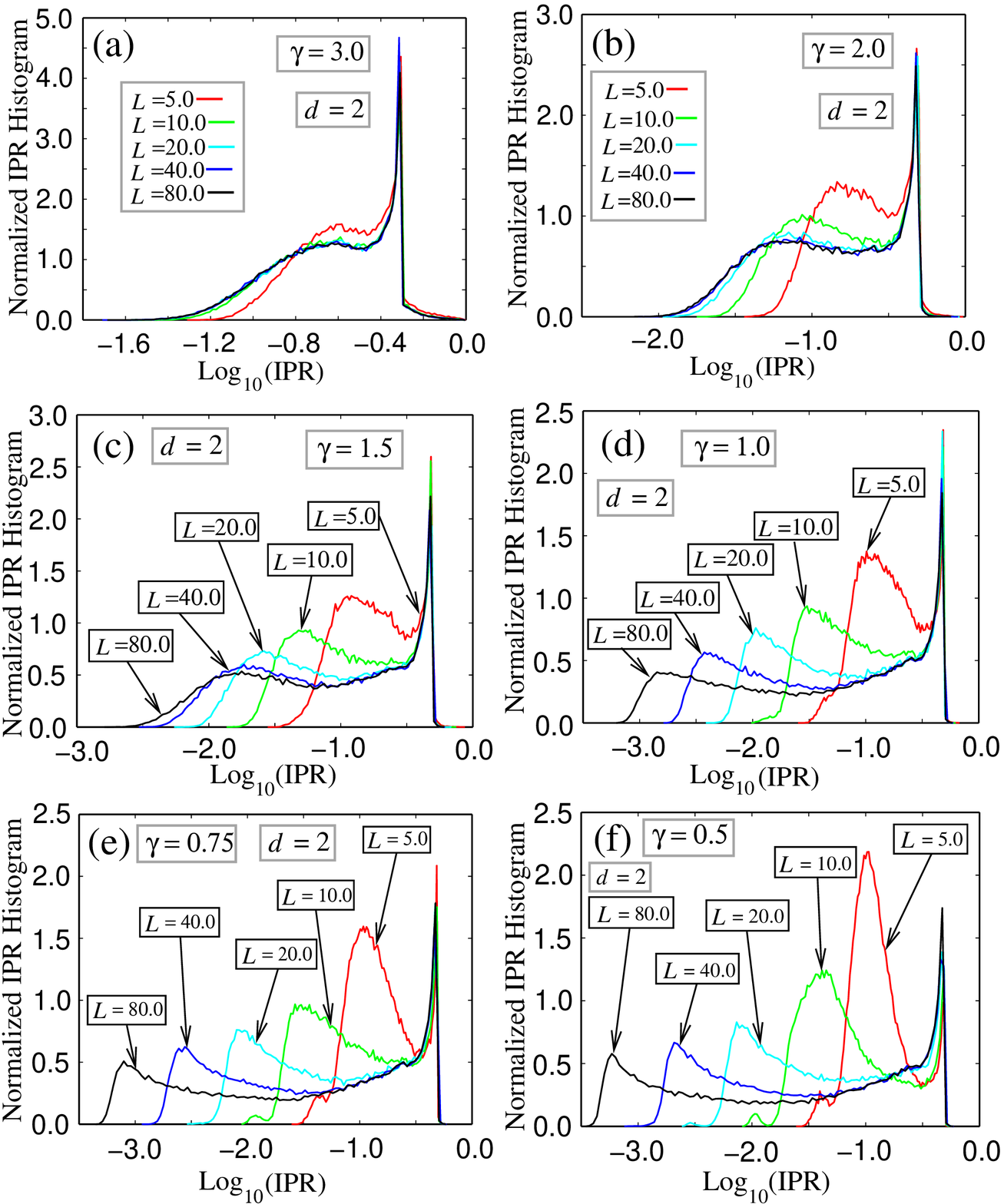}
\caption{\label{fig:Fig4} (Color Online) Inverse Participation Ratio profiles graphed for 
various system sizes $L$ for hopping integral decay rates
decay rates ranging from $\gamma = 3.0$ in panel (a) to $\gamma = 0.5$ in panel (f) for 2D systems.}
\end{figure}

\begin{figure}
\includegraphics[width=.49\textwidth]{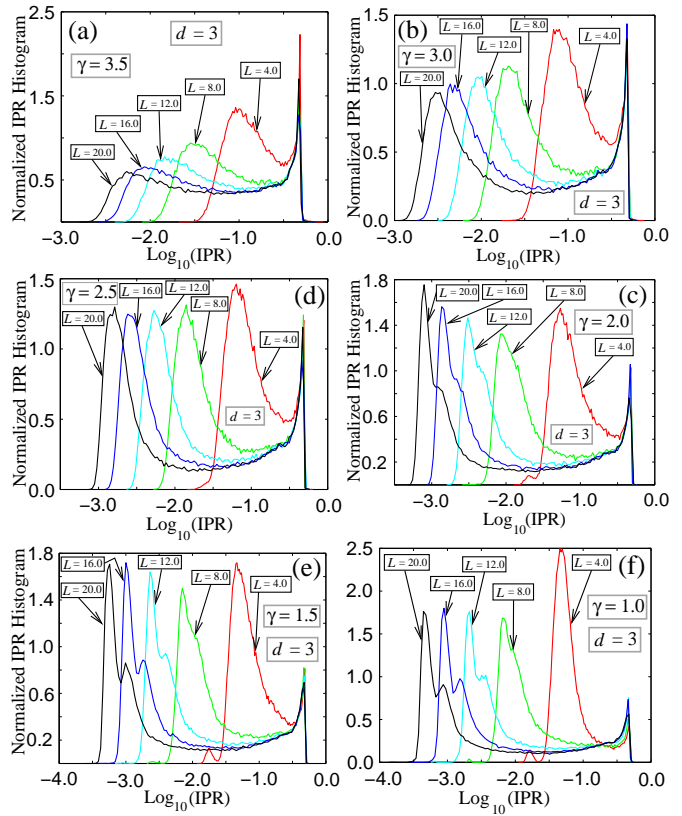}
\caption{\label{fig:Fig5} (Color Online) Inverse Participation Ratio profiles graphed 
for various system size $L$ for hopping integral 
decay rates $\gamma$ ranging from $\gamma = 3.5$ in panel (a) to $\gamma = 1.0$ in panel (f) for 3D systems.}
\end{figure}

The Inverse Participation Ratio (IPR) is a compact single parameter gauging the degree 
to which electronic states are localized or extended, with the IPR tending to zero for $L \rightarrow \infty$
for bulk extended states.  While our aim in this work is to extrapolate to the bulk limit in a quantitative 
fashion, information may also be gleaned at a qualitative level when participation ratio curves are 
juxtaposed for a range of system sizes.  Since the IPR may vary by several orders of 
magnitude over the full gamut of system sizes $L$ under consideration, it is often more prudent to 
exhibit $\log_{10} (\mathrm{IPR})$ in lieu of the raw participation ratios. 

The IPR histograms are created by dividing the interval along the $\log_{10}(\mathrm{IPR})$ abscissa 
into a suitable number of bins. 
With the availability of $10^{5}$ participation ratios, the choice of 200 
partitions provides a reasonable measure of resolution while keeping statistical 
fluctuations at reasonable levels.  IPR distributions are shown in Fig.~\ref{fig:Fig3}, 
Fig.~\ref{fig:Fig4}, and Fig.~\ref{fig:Fig5} for $D=1$, $D=2$, and $D=3$ respectively.

In the 1D case, the system sizes progress geometrically, doubling from $L = 100$ to 
$L = 6400$.  With successive 
doublings of $L$, the IPR histograms invariably converge and cease to 
evolve with increases in the system size, a 
characteristic consistent with the localization of all wave functions in the thermodynamic limit.
When converged in $L$ to an IPR distribution appropriate to the bulk limit, 
histograms are dominated by cusp-like peaks near the upper limit of the IPR range.
The latter characteristic is a hallmark particular to 1D systems
and is evident whether the hopping integral length scale $l$ is large or small relative to the 
separation between sites.

The interpretation from the $D = 1$ participation ratio distributions that all wave functions
are localized is in a sense not surprising.
With the theoretical 
framework of Anderson localization having been introduced more than 50 years ago~\cite{Anderson}, 
a significant body of work (both in experiment and theoretical calculations) has examined 
the tendency for random potentials to localize electronic states very effectively
in one dimensional systems, even for weak random potentials. 
Moreover, the availability of cold atom traps with coherent quantum states 
where the underlying one dimensional potential may be tailored in a variety of ways has 
made possible the study of localization properties of 1D systems in a controlled 
manner.  In this vein, a direct experimental observation of localization has recently been 
achieved in a Bose-Einstein condensate with the random (diagonal) potential 
set up by a laser speckle~\cite{natbilly} with results in accord with theoretical 
descriptions~\cite{simspeckle}.  Bichromatic aperiodic potentials are not
purely random uncorrelated disorder, but nonetheless have been found in experiment and 
theoretical analysis~\cite{Biddle1,Biddle2,Biddle3} to be very effective in localizing quantum states.
In this work, we show using 
finite size scaling analysis that off-diagonal disorder inherent in one dimensional 
amorphous systems leads to the localization of all electronic states in the absence of 
random site potentials.

For $D = 2$, the evolution of the participation ratio with increasing $L$ depends on the hopping 
integral decay length $l$.
IPR histograms for various $\gamma$ values are shown in the six panels of Fig.~\ref{fig:Fig4};    
as in the 1D case, successive values of $L$ differ by a factor of 2, and participation ratio distributions 
for $L = \left \{ 5, 10, 20, 40, 80  \right \}$ appear together on the same graph. 
For short hopping ranges $l = \gamma^{-1}$, there are  
maxima in the high IPR regime which subsume a large share of 
the total statistical weight; apart from 
small vacillations, the distributions have a unimodal envelope,
as seen in IPR histograms calculated for one dimensional systems. 
Moreover, the convergence of histograms with respect to $L$ is an 
indication of the localization of all states in the thermodynamic limit. 

For longer decay lengths $l$ (i.e. for $\gamma \leq 1.5$) participation ratio distributions obtained for 
2D amorphous systems lack the unimodal sharply peaked structure 
seen for small $l$ and unlike
the $l \ll 1$ counterparts, seem not 
to converge with respect to increases in $L$.   
Instead, histograms become bimodal as statistical weight is transferred to the left,
toward the lower IPR regime.

The migration of statistical weight toward lower IPR values and 
more extended character, is indicative of the possible existence of extended states,
and the effect is particular striking for the smallest decay rates $\gamma = 0.75$ 
and $\gamma = 0.5$.  The systematic transfer with increasing $L$ of probability density to smaller IPR 
values is manifest as a steady leftward shift of the trailing (low IPR) edge and of the 
histogram with the nearby maximum also born leftward.  In fact, the shift both of the peak and 
the left-most front is constant in magnitude each time the system size $L$ is doubled; since the 
abscissa is $\log_{10} (\mathrm{IPR} )$, the latter trend corresponds to a power law 
scaling $L^{-\beta}$ for the left trailing edge and low IPR peak.

Participation ratio histograms for 3D systems are displayed in Fig.~\ref{fig:Fig5}. 
As in the large $l$ limit for 2D systems, a portion of the participation ratio 
statistical weight shifts systematically to lower IPR values.  The size of the emerging peak 
and the amount of probability density which migrates leftward
increases with $l$, even as the cusp-shaped maximum in the low IPR regime decreases in 
amplitude and overall statistical weight.  

A noteworthy feature for $l \leq 1$ is the robustness of 
the probability density contained in the leftward shifting peak.  That the packet 
moves systematically toward lower IPR values without leaving behind any statistical 
weight is compatible with the existence of a finite fraction of extended states in the 
bulk limit.  Thus, for 3D amorphous systems, there is a dichotomy in the 
way the distribution changes, 
where the leftward shifting probability density corresponding to 
extended wave functions contrasts with the localized states 
encompassed in the high IPR region of the histogram, which ceases to evolve      
with increasing $L$.  With increasing tunneling matrix element range $l$, the 
balance shifts in favor of the extended states as more and more statistical weight is swept into     
the peak moving toward lower IPR values.  Nonetheless, as embodied in the 
part of the distribution which does not change with increasing 
$L$, a finite fraction of the wave functions 
are localized even for very small values of the decay rate $\gamma$.

\section{The Channel Averaged Participation Ratio}

\begin{figure}
\includegraphics[width=.49\textwidth]{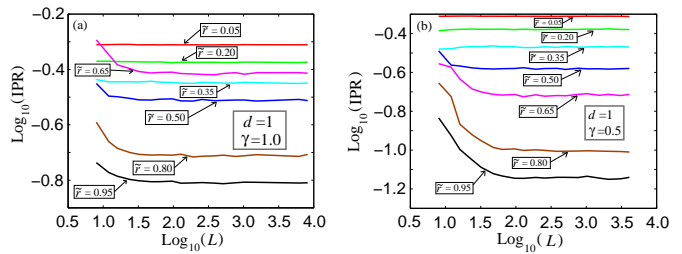}
\caption{\label{fig:Fig6} (Color Online) Log-Log graphs of channel averaged 
IPR versus $L$ for 1D systems in the case of a 100 channel scheme.
Panel (a) and panel (b) correspond to hopping integral decay rates 
$\gamma = 1.0$ and $\gamma = 0.5$ respectively.} 
\end{figure}

\begin{figure}
\includegraphics[width=.49\textwidth]{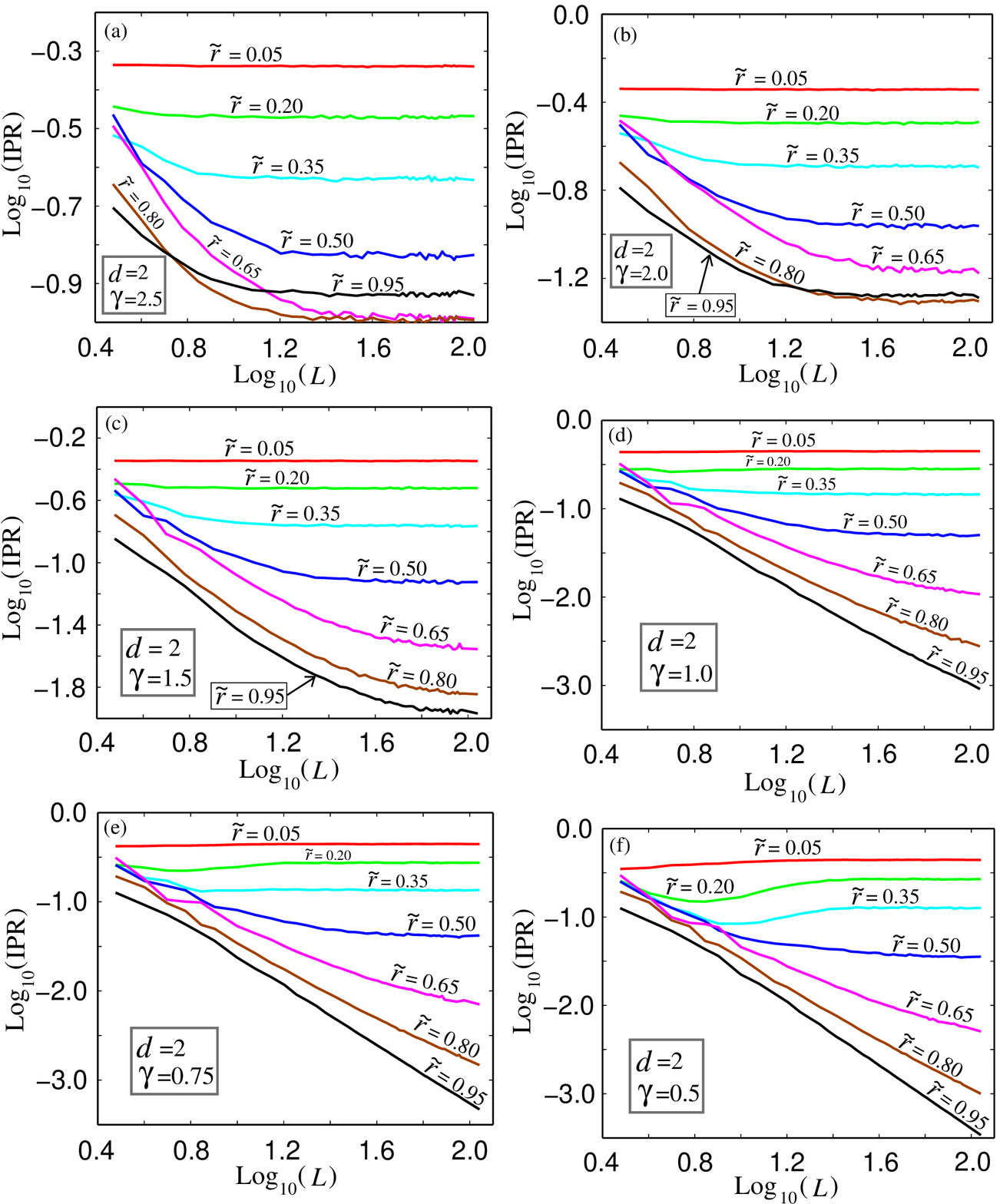}
\caption{\label{fig:Fig7} (Color Online) Log-Log graphs of channel averaged
IPR versus $L$ for 2D systems in the case of a 100 channel scheme with
hopping integral decay rates ranging from $\gamma = 2.5$ in panel (a) 
to $\gamma = 0.5$ in panel (f).}  

\end{figure}

\begin{figure}
\includegraphics[width=.49\textwidth]{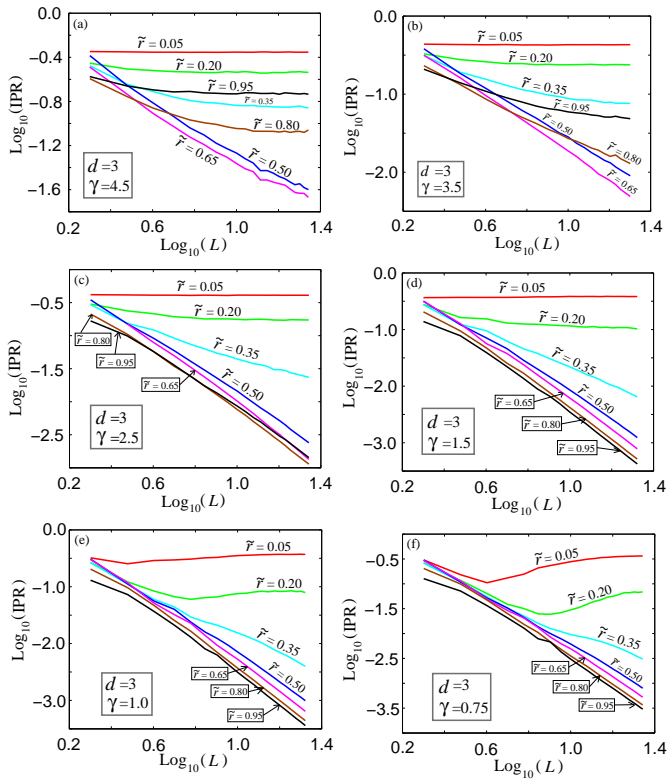}
\caption{\label{fig:Fig8} (Color Online) Log-Log graphs of channel 
averaged IPR versus $L$ for 3D systems in the case of a 100 channel scheme with
hopping integral decay rates ranging from $\gamma = 4.5$ in panel (a) 
to $\gamma = 0.75$ in panel (f).}
\end{figure}

With long-range positional order absent in the amorphous systems we examine, the eigenstate energy is the only good 
quantum number available, and in this work we show that the energy eigenvalue is a unique determinant  
as to whether electronic states are either extended or localized for a particular energy
with the simultaneous presence of localized and extended states ruled out as a possibility.

Determining if a wave function $\psi$ is localized or extended entails
calculating participation ratio statistics for many $L$ values and 
using finite size scaling to access the $L \rightarrow \infty$ limit, with a vanishing IPR in the 
bulk limit a hallmark of extended states.  The random character 
of the disorder precludes the study of the evolution of individual states with increasing 
system size, and instead we must analyze aggregates of wave functions across a range of 
system sizes.  Electronic states are parameterized by energy eigenvalues, and one possible choice is 
to partition the states into channels centered about uniformly spaced energies $E^{'}$ 
which encompass an energy interval $\delta E$ narrow enough to 
capture information specific to wave functions with energies very close to $E^{'}$, but 
broad enough to suppress statistical fluctuations.

The DOS statistical distributions shown in Fig.~\ref{fig:Fig2} are sharply peaked, with 
a rapid decrease in the probability density away from the maximum.  
In a practical sense, the DOS heterogeneity poses a challenge for a scheme where the 
global energy range $\Delta E$ between the ground state energy $E_{\mathrm{min}}$ and the highest excited 
state energy $E_{\mathrm{max}}$ is partitioned into small intervals $\delta E$ of uniform 
size; statistical fluctuations will plague channels far from the DOS maximum where the statistical 
weight for eigenstates is sharply reduced.

In lieu of energy, to circumvent the problem of non-uniform statistics, 
states are labeled with the normalized energy eigenvalue rank $\tilde{r}$.
For finite systems, the rank number is assigned by calculating energy eigenvalues and corresponding
wave functions for a large number of configurations of disorder.  The normalized rank number is $\tilde{r} \equiv 
r/N$ where $N$ is the total number of states and $r$ is the global eigenvalue rank 
within the large aggregate.  We find $n = 100$ yields 
channel widths $\delta \tilde{r}$ sufficiently narrow for channel averages to be representative of 
the normalized rank $\tilde{r}$ at the center of the channel, 
yet broad enough to provide sufficient statistics for analysis; a parallel calculation 
for $n = 50$ yields results in quantitative agreement with the $n = 100$ scheme, direct verification 
that $n = 100$ is large enough to avoid systematic admixture effects from the finite channel width $\delta \tilde{r}$.

Channel averaged IPR results appear in Fig.~\ref{fig:Fig6}, Fig.~\ref{fig:Fig7}, and Fig.~\ref{fig:Fig8}
for 1D, 2D, and 3D systems, where the horizontal axis is $\log_{10}(L)$ for each case.
The participation ratio results for the one dimensional systems are in a sense most
 readily interpreted.  IPR curves initially decrease for small system sizes, but quickly 
level out and approach asymptotically finite values
even for the small decay constant $\gamma = 0.5$.  The latter phenomenon is consistent with the 
convergence of the global IPR probability distribution with increasing $L$, interpreted as a sign 
that all wave functions are localized in one dimensional systems irrespective of the hopping decay 
parameter $\gamma$. 

For $D = 2$, there is a bifurcation in the way the IPR channel averages 
vary with increasing $L$
with the precise behavior determined by the hopping range $l = \gamma^{-1}$.  
For large $\gamma$ (i.e. especially for $\gamma = 2.0$ and 
$\gamma = 2.5$), the IPR traces level out and approach finite    
asymptotic bulk values corresponding to localized states, much as occurs for 1D amorphous systems.  On the other hand, 
in the case of smaller $\gamma$, where there is a greater likelihood that states with extended character are supported,  
it is not definitively conclusive that the participation ratio 
curves approach asymptotically finite values for very large system sizes.  Broadly speaking,  
for $D = 2$ IPR traces are most likely to level out and tend to a finite value for smaller 
$\tilde{r}$, or for energies near the ground state.  For $\gamma \leq 1.0$, there 
are channel average curves which in principal may keep a finite negative slope as $L \rightarrow \infty$, which 
on a logarithmic scale is tantamount to a vanishing IPR in the thermodynamic limit.

Notwithstanding the persistently downward slope, an unambiguous 
determination that the monotonically decreasing channel averages represent 
extended states is hampered by the upward concavity in 
most of the IPR curves, where except for the 
very highest $\tilde{r}$ (i.e. for $\tilde{r} = 0.95$ where the concavity appears to be neutral),  
there is a progressive reduction in the magnitude of the downward slope which could eventually 
cause the channel averaged participation ratio to level out at a finite value.
To determine in a rigorous way if the participation ratios vanish for  $L \rightarrow \infty$
or instead approach a finite value, a finite size scaling analysis, described in Section V, is needed.

As in the case of the 2D systems, for $D = 3$, there are in a broad sense two ways in which 
IPR curves scale with $L$.  For very large decay rates (e.g. $\gamma = 4.5$), the participation ratio 
channel averages seem to level out and tend to finite values. On the other hand
for smaller $\gamma$, instead of the upward concavity seen in the case of two 
dimensional systems, many of the curves are concave downward with the rate of decrease of the 
channel averages increasing with $L$.  With the downward slope becoming greater instead of showing  
signs of faltering as in $D=2$, it is possible to conclude without further analysis that the 
participation ratio tends to zero in the bulk limit.

\begin{figure}
\includegraphics[width=.49\textwidth]{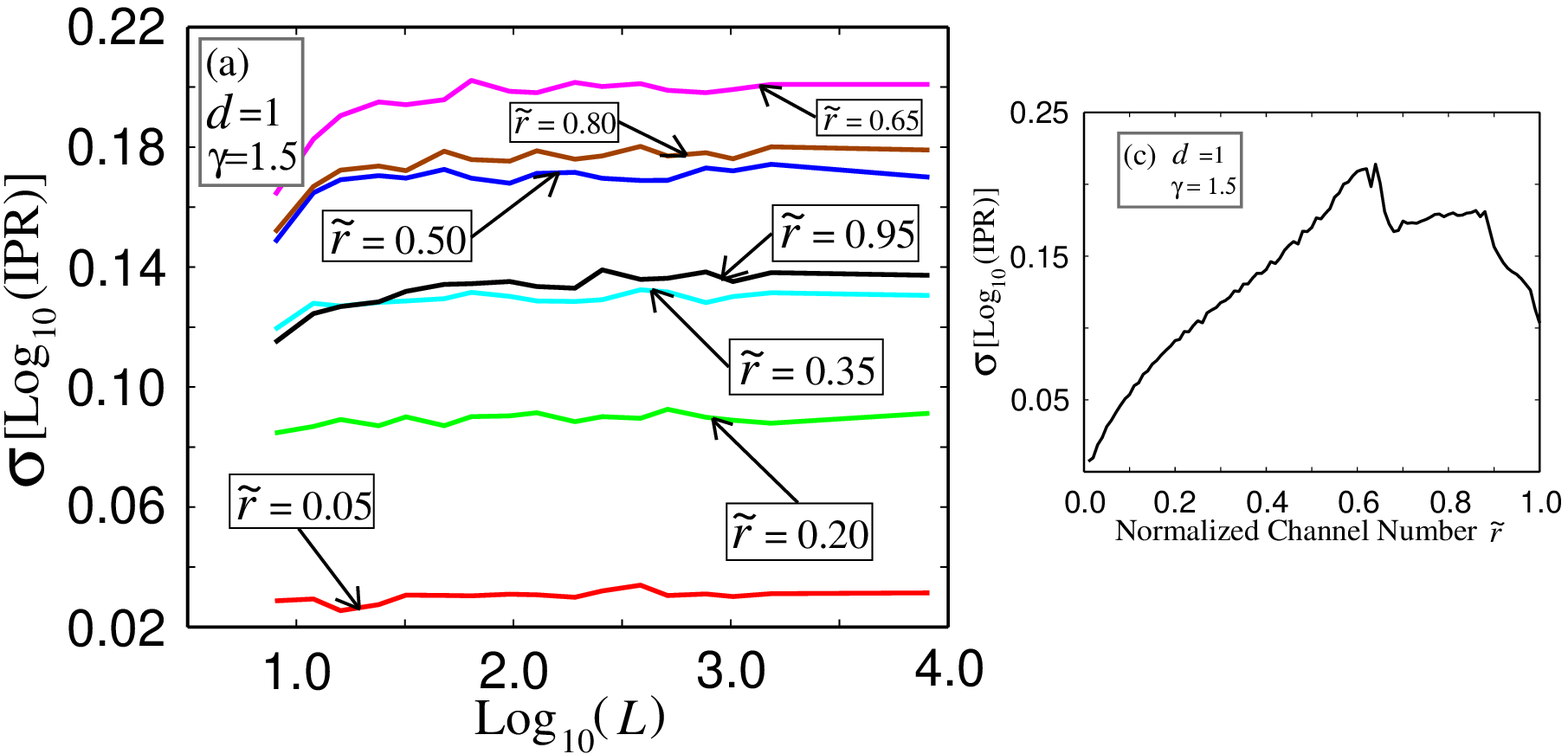}
\includegraphics[width=.49\textwidth]{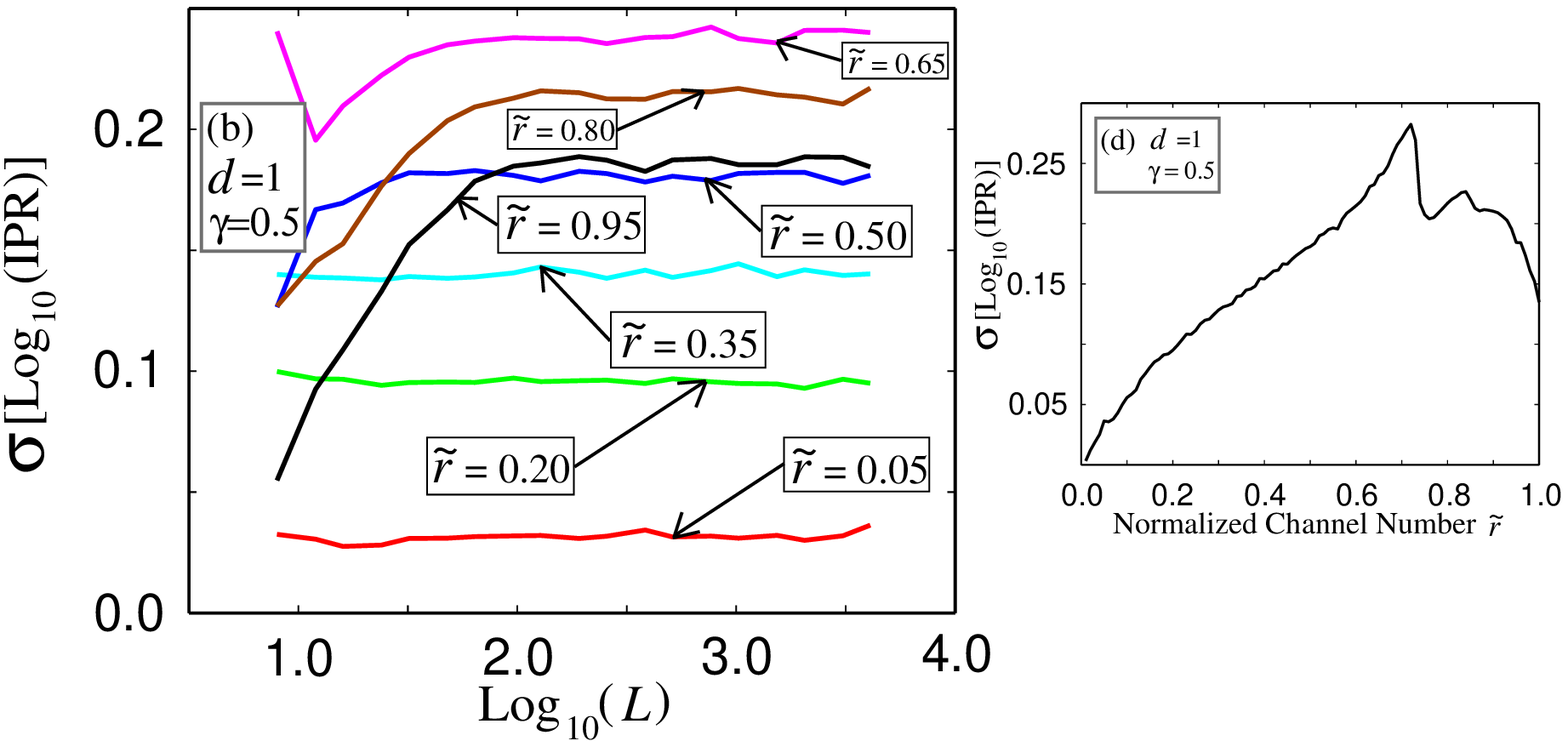}
\caption{\label{fig:Fig9} (Color Online) Intra-channel standard deviations of $\log_{10}(\mathrm{IPR})$ 
plotted for the $D = 1$ case.  Graphs in panels (a) and (c) correspond to relatively large 
$\gamma = 1.5$, while plots in panels (b) and (d) are calculated for a more gradual hopping integral 
decay, $\gamma = 0.5$.  Graphs on the left show $\sigma$ versus $\log_{10}(L)$ for various   
normalized channel numbers $\tilde{r}$, while the rightmost panels are plots of 
$\sigma[\log_{10}(\mathrm{IPR}]$ with respect to the rank $\tilde{r}$ for large $L$.} 
\end{figure}

\begin{figure}
\includegraphics[width=.49\textwidth]{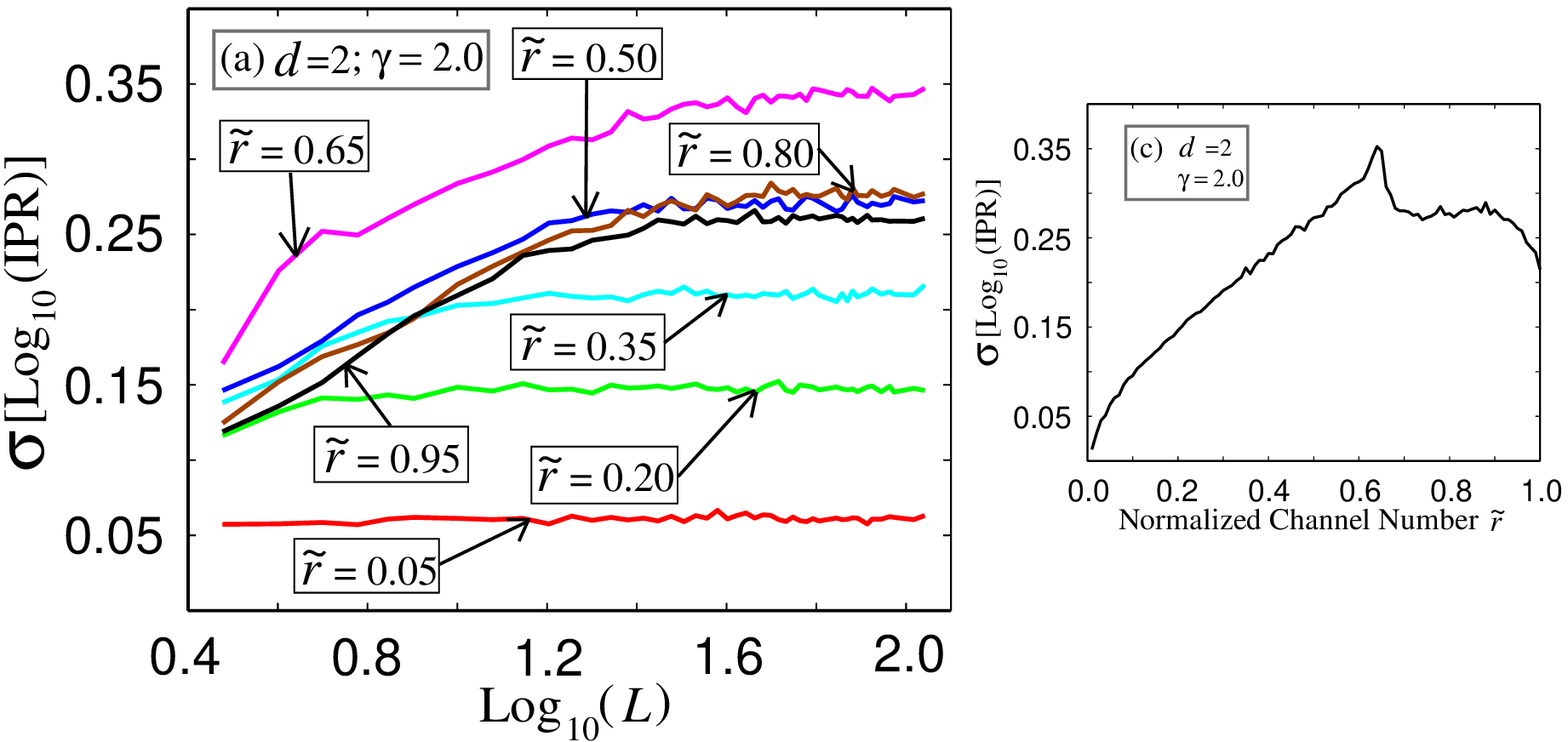}
\includegraphics[width=.49\textwidth]{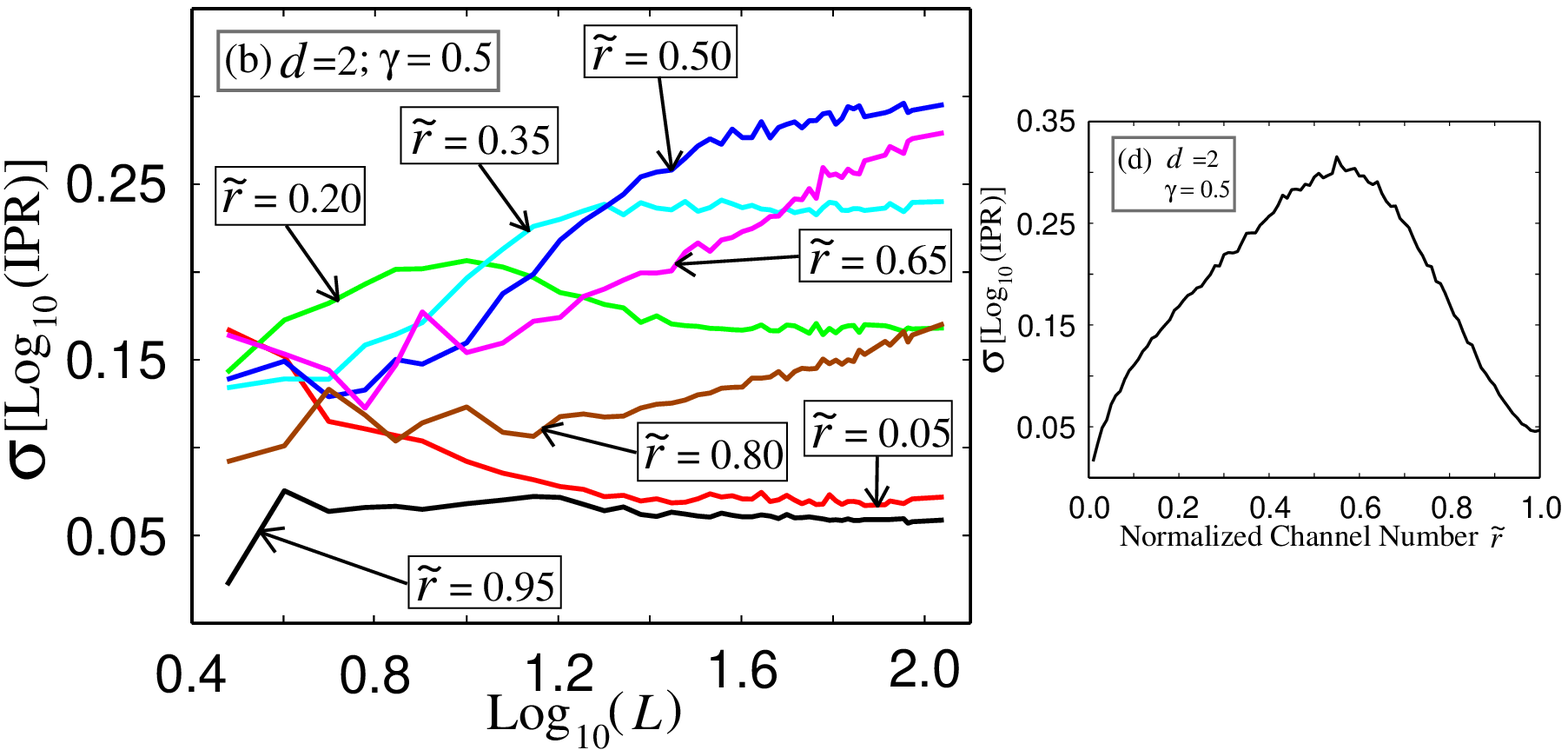}
\caption{\label{fig:Fig10} (Color Online) Intra-channel standard deviations of $\log_{10}(\mathrm{IPR})$ 
plotted for the $D = 2$ case.  Graphs in panels (a) and (c) correspond to relatively large 
$\gamma = 2.0$, while plots in panels (b) and (d) are calculated for a more gradual hopping integral 
decay, $\gamma = 0.5$.  Graphs on the left show $\sigma$ versus $\log_{10}(L)$ for various    
normalized channel numbers $\tilde{r}$, while the rightmost panels are plots of 
$\sigma[\log_{10}(\mathrm{IPR}]$ with respect to the rank $\tilde{r}$ for large $L$.}  
\end{figure}

\begin{figure}
\includegraphics[width=.49\textwidth]{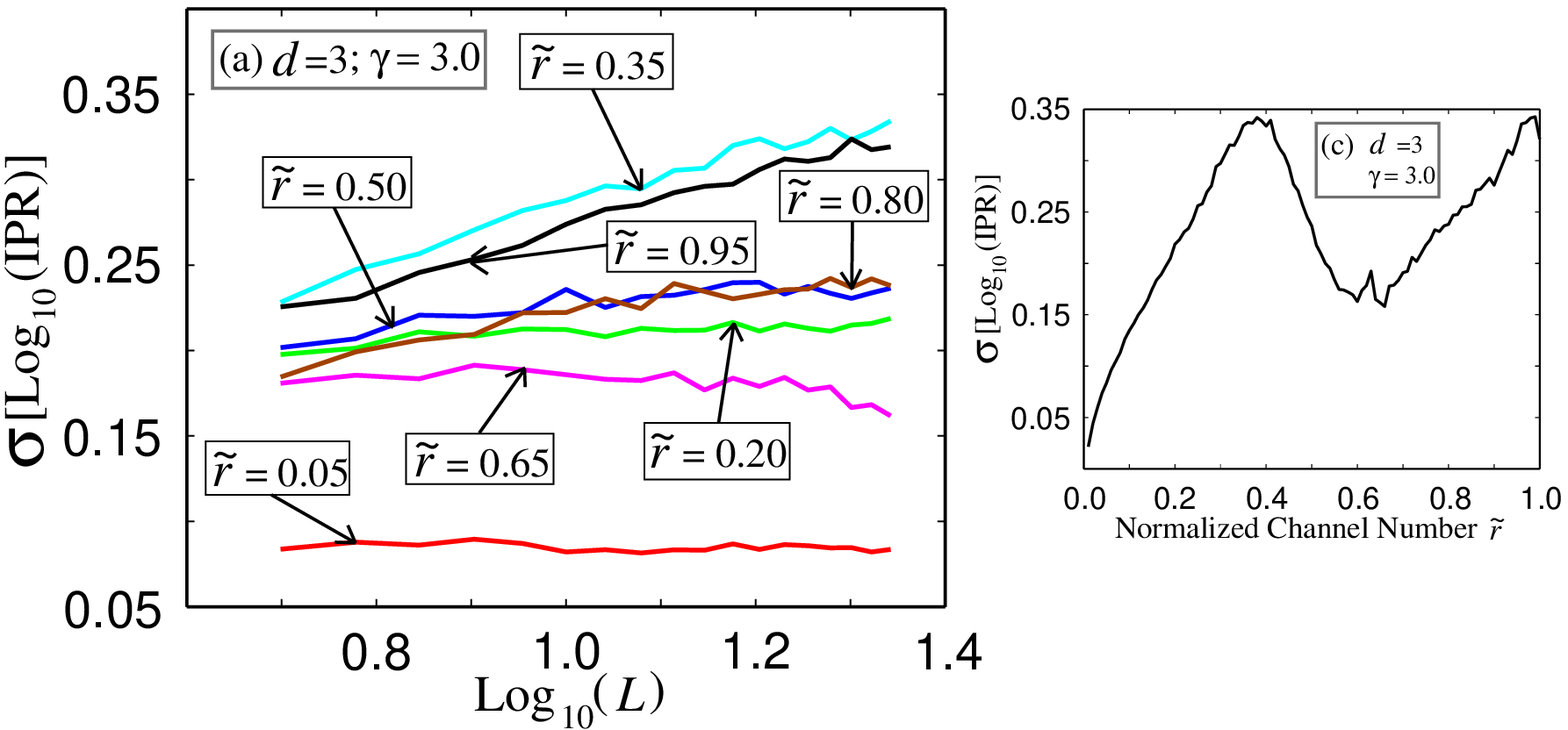}
\includegraphics[width=.49\textwidth]{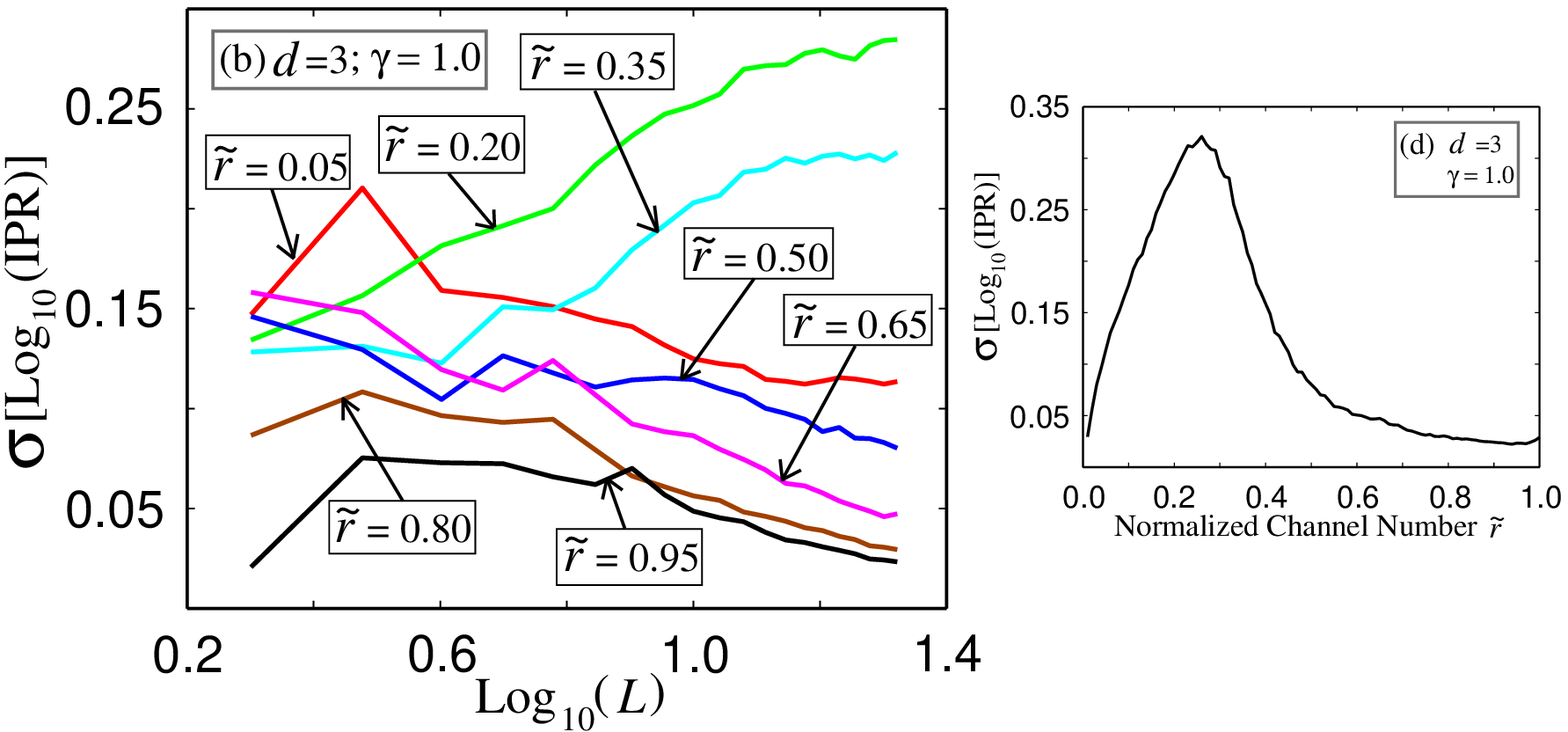}
\caption{\label{fig:Fig11} (Color Online) Intra-channel standard deviations of $\log_{10}(\mathrm{IPR})$
plotted for the $D = 3$ case.  Graphs in panels (a) and (c) correspond to relatively large
$\gamma = 3.0$, while plots in panels (b) and (d) are calculated for a more gradual hopping integral
decay, $\gamma = 1.0$.  Graphs on the left show $\sigma$ versus $\log_{10}(L)$ for various
normalized channel numbers $\tilde{r}$, while the rightmost panels are plots of
$\sigma[\log_{10}(\mathrm{IPR}]$ with respect to the rank $\tilde{r}$ for large $L$.}  
\end{figure}

With channel averages only providing the mean participation ratio, it is important to be certain the 
IPR values obtained in this manner represent the characteristics of all of the states encompassed in a channel.   
The most vital task in this vein is to 
rule out the possibility of the coexistence of localized and extended states for a specific 
channel index $\tilde{r}$ (or energy $E$). 
To tackle this question, we calculate the standard deviation $\sigma$ of $\log_{10} (\mathrm{IPR})$ across
a channel with
results appearing Fig.~\ref{fig:Fig9} for $D = 1$, Fig.~\ref{fig:Fig10} for $D = 2$, and Fig.~\ref{fig:Fig11} for 
$D = 3$.  
With increasing $L$, the quantity $\sigma[\log_{10} (\mathrm{IPR} )]$ will either diverge or tend to a finite value.
While a divergence indicates a dual presence of extended and localized states (coexistence), a standard deviation
of participation ratio logarithms
which tends to a finite value for $L \rightarrow \infty$ is an indication that there are either localized or extended wave functions, 
but not both (exclusivity). 

If only localized states are present, the 
the standard deviation will converge to a limiting value ($\sigma$ is in general finite even in the thermodynamic 
limit since some wave functions are more spread out than others due to disorder fluctuations) 
and become a bulk characteristic when $L \gg \xi$, where $\xi$ is the largest decay length scale of the broadest states.
On the other hand, for a suite of purely extended states, the IPR for 
each wave function will tend to zero with a concomitant divergence of 
$\log_{10}(\mathrm{IPR})$.  Although the magnitude of the ensemble average 
$\langle \log_{10} (\mathrm{IPR}) \rangle$, which is negative in sign,
 grows without bound for increasing $L$ for a rank $\tilde{r}$ or energy $E$ supporting extended states,
it is nevertheless not clear that $\sigma$ diverges since the standard deviation measures the spread in the 
participation ratios, not their magnitude.  In 
the global IPR histograms shown in Fig.~\ref{fig:Fig4} and 
Fig.~\ref{fig:Fig5} for $D = 2$ and $D = 3$, the fact that
packets of probability density which migrate lower $\log_{10}(\mathrm{IPR})$  
with increasing $L$ tend to maintain their shape and profile even as the participation ratio decreases by several 
orders of magnitude suggests $\sigma$ may remain finite even as the average $\langle \log_{10} (\mathrm{IPR}) \rangle$ diverges.

In the coexistence scenario, with a simultaneous presence of extended and localized wave functions, there will be a divergence 
in $\sigma[ \log_{10} (\mathrm{IPR})]$ since the packet of probability density for localized states remains fixed with 
increasing $L$ while the peak corresponding to states with extended character is conveyed toward more negative 
$\log_{10} (\mathrm{IPR})$. Since the standard deviation provides a measure of the increasing separation of the two peaks, 
$\sigma [ \log_{10} ( \mathrm{IPR} )]$ must diverge as $L \rightarrow \infty$ if localized and extended electronic 
states are present for the same energy $E$ or rank $\tilde{r}$.

To determine if for any combination of system parameters $\sigma$ converges 
to a finite value (exclusivity) or diverges (coexistence),
we plot $\sigma[\log_{10} (\mathrm{IPR} )]$ versus $\log_{10}(L)$  
in the left panels of the graphs in Figs.~\ref{fig:Fig9} 
Figs.~\ref{fig:Fig10}, and Figs.~\ref{fig:Fig11}. 
In addition, the right panels show  
$\sigma$ with respect to $\tilde{r}$ for large $L$ to provide a glimpse of $\sigma[\log_{10}(\mathrm{IPR})]$
for large $L$.  The $\sigma$ curves for the cases $D = 1$, and $D = 2$, and $D = 3$ are concave downward in the 
large $L$ regime, and seem to level out and tend to finite values.  In addition, for the largest system size 
considered in this work and plotted in the right panels of 
Fig.~\ref{fig:Fig9}, Fig.~\ref{fig:Fig10}, and Fig.~\ref{fig:Fig11},
$\sigma[\log_{10}(\mathrm{IPR})]$ are low in magnitude, and do not exceed $\sigma[\log_{10}(\mathrm{IPR})] = 0.35$
for which the spread of the participation ratios is at most a factor of two.

\section{Participation Ratios in the Thermodynamic Limit}     

The primary aim in the finite size scaling 
analysis is to calculate $\mathrm{IPR}_{0}$, the channel averaged participation ratio in the thermodynamic limit.
For the IPR dependence for moderate to large $L$, we use a power law formula,
$\mathrm{IPR}(L) = \mathrm{IPR}_{0} + \alpha_{1} L^{-\beta} + \alpha_{2} L^{-\delta}$ where 
$\mathrm{IPR}_{0}$ is the participation ratio in the bulk limit, $\beta$ is the leading order scaling exponent, 
and $\delta$ is the exponent for the next to 
leading order contribution to scaling; $\alpha_{1}$ and $\alpha_{2}$ are amplitudes.  
We obtain the parameters $\mathrm{IPR}_{0}$, $\beta$, $\delta$, $\alpha_{1}$, and $\alpha_{2}$ 
in a nonlinear least squares calculation by
minimizing the sum of the square of the relative differences $\Delta_{\mathrm{LSF}}
 \equiv \frac{1}{m}[\sum_{i=1}^{m} (\frac{\mathrm{IPR}^{\mathrm{CA}}(L_{i})-
\mathrm{IPR}^{\mathrm{LSF}}(L_{i})}{\mathrm{IPR}^{\mathrm{CA}}(L_{i})})^{2}]^{1/2}$
(i.e. with $L_{1}$ and $L_{m}$ the smallest and largest systems examined, respectively)
of the data gleaned from the IPR  
channel averages $\mathrm{IPR}^{\mathrm{CA}}(L)$ 
and the theoretical scaling expression $\mathrm{IPR}^{\mathrm{LSF}}(L)$.  To find the optimal fit, 
we use a stochastic algorithm with the quantity $\Delta_{\mathrm{LSF}}$ treated as an ``energy'' to be minimized 
by randomly perturbing $\mathrm{IPR}_{0}$, amplitudes $\alpha_{1}$ and $\alpha_{2}$, and exponents
$\beta$ and $\delta$.   Only Monte Carlo moves which 
decrease $\Delta_{\mathrm{LSF}}$ (and hence incrementally improve the fit) are accepted, and the $\mathrm{IPR}(L)$ parameters are suitably converged 
after $4 \times 10^{5}$ attempts to shift the five unknown parameters in $\mathrm{IPR}(L)$ in a stochastic
fashion.

\begin{figure}
\includegraphics[width=.49\textwidth]{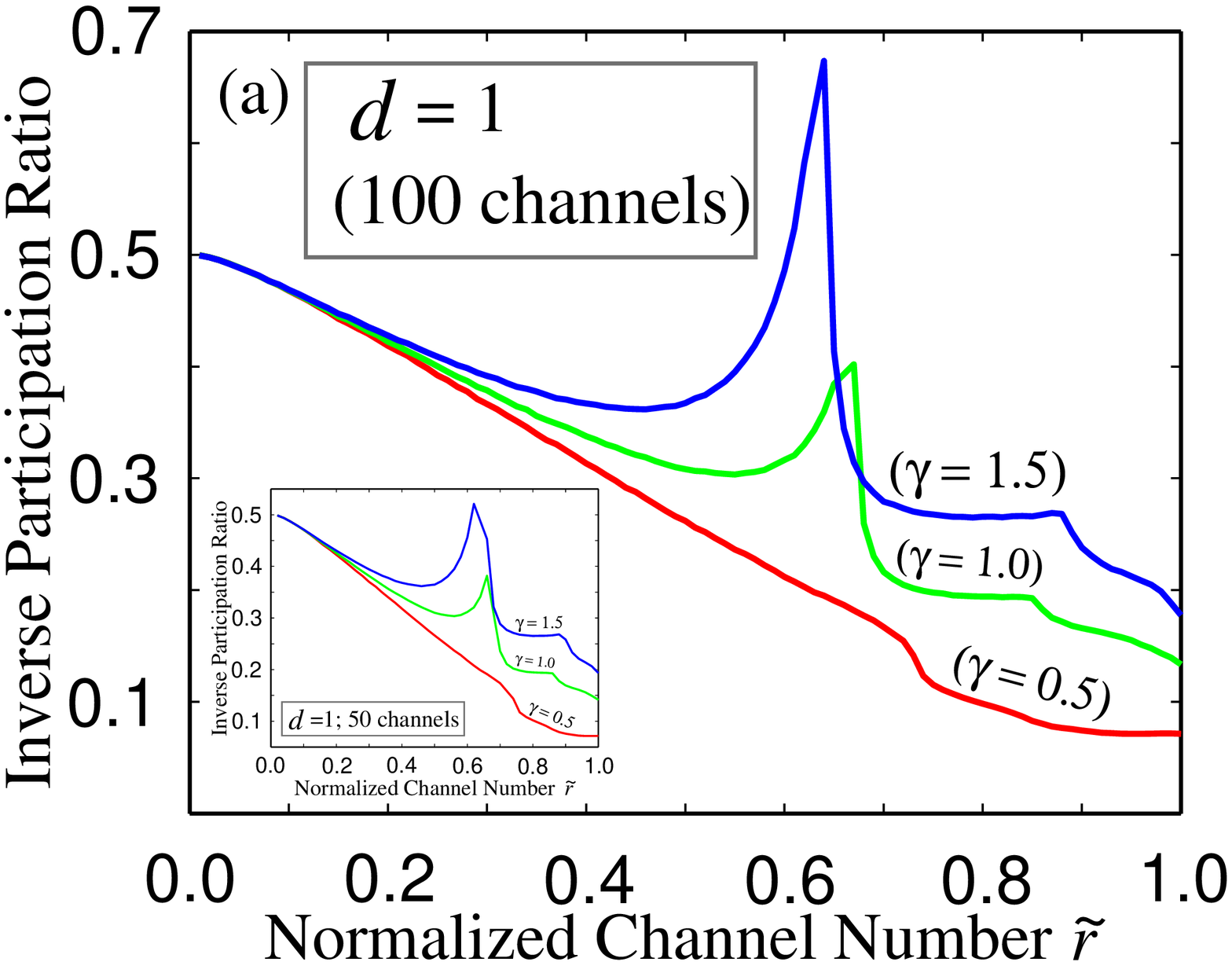}
\includegraphics[width=.49\textwidth]{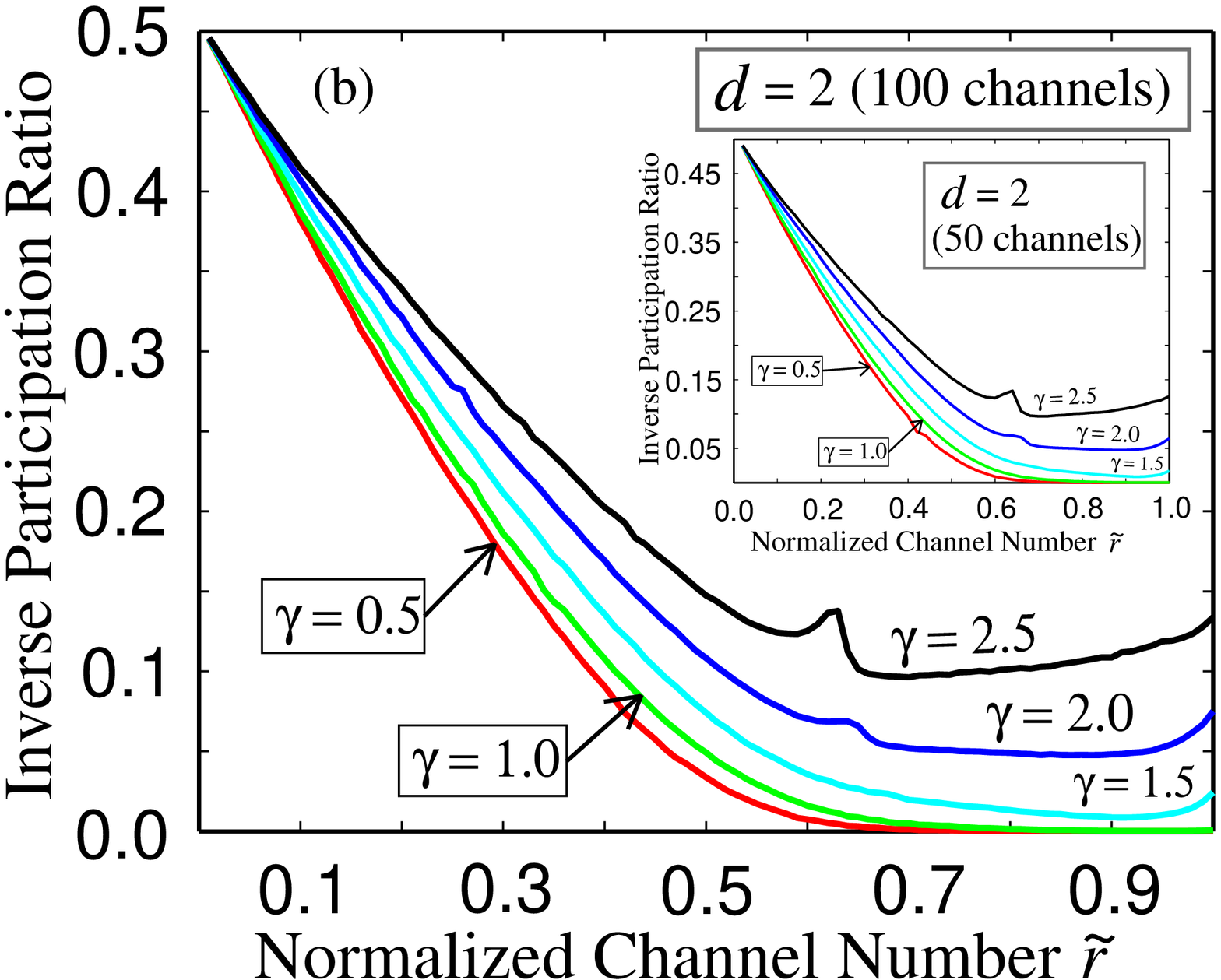}
\includegraphics[width=.49\textwidth]{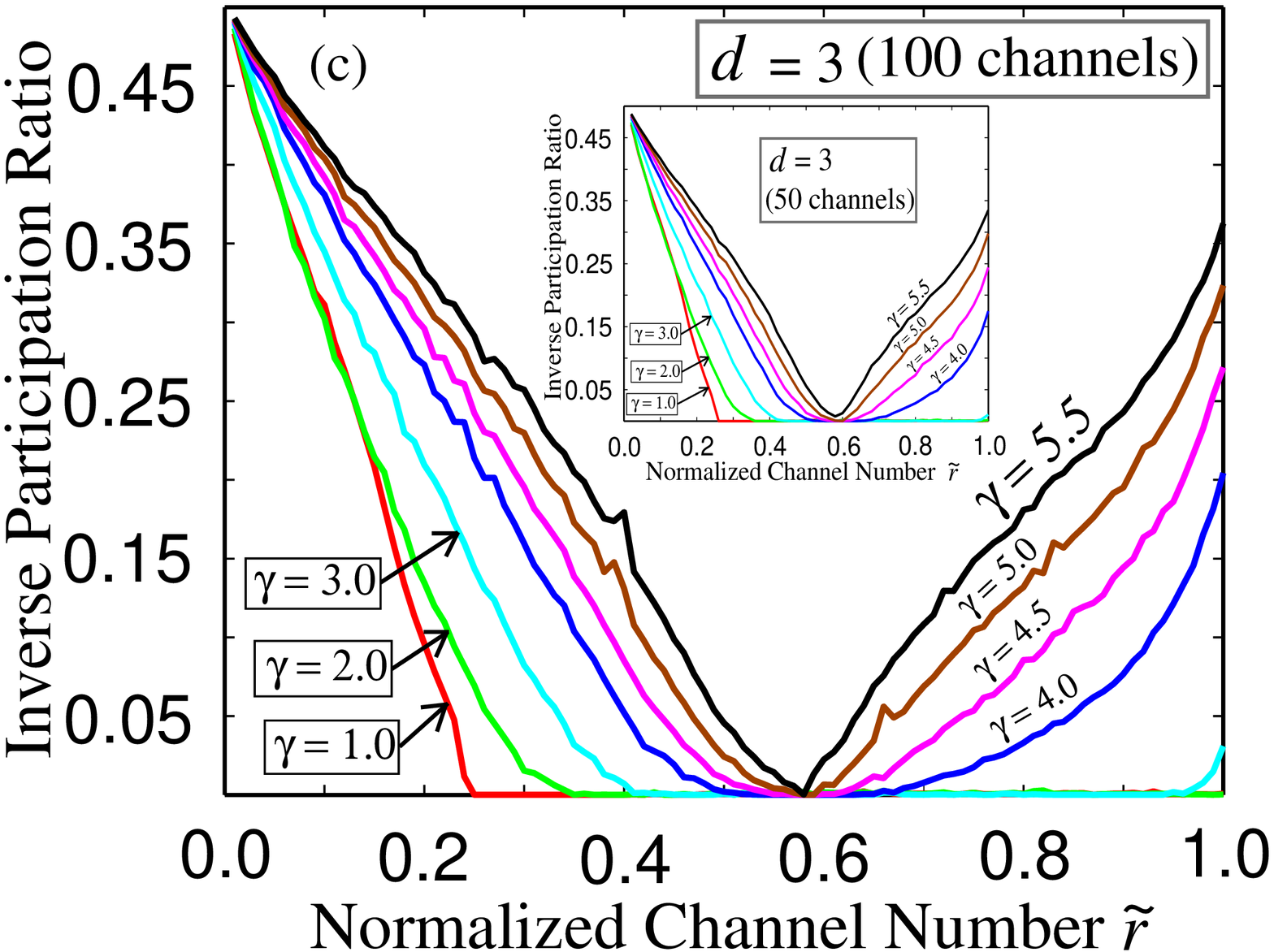}
\caption{\label{fig:Fig12} (Color Online) Inverse Participation Ratio profiles graphed for strong
decay rates $\gamma$ [ranging from $\gamma = 3.5$ in panel (a) to $\gamma = 2.0$ in panel (d)] for 1D systems.}
\end{figure}

Participation ratios extrapolated to the bulk limit are displayed for
$D = 1$, $D = 2$, and $D = 3$ in Fig.~\ref{fig:Fig12}.
The main graphs are calculated for 100 channel partitions, whereas the insets show results 
for a 50 channel scheme.  The good agreement among the 100 and 50 channel bulk IPR values  indicates
convergence of the bulk IPR values with respect to the number of channels.  

Bulk limit Participation ratios for 1D systems are plotted in panel (a) 
of Fig.~\ref{fig:Fig12}.  Notwithstanding a nonmonotonic variation of IPR in the normalized
channel number $\tilde{r}$ the $D = 1$ results are finite in all cases, 
dipping only slightly below $\mathrm{IPR}_{0} = 0.1$ even for hopping integral decay rates as low as $\gamma = 0.5$, 
and thus are 
indicative of localization of all wave functions irrespective of the size of the hopping range 
$l$.

Extrapolated IPR results for $D = 2$ shown in panel (b) of Fig.~\ref{fig:Fig12}
are non-monotonic, generally decreasing precipitously near $\tilde{r} = 0$ and ultimately recovering in
the vicinity of $\tilde{r} = 1$.  With increasing $l$, the participation ratio trough becomes broader and deeper, 
while the recovery to higher IPR values is muted.  Eventually, for $\gamma \leq 1$, the minimum reaches the 
horizontal axis, where $\mathrm{IPR}_{0} = 0$, and there is no return to higher participation ratios near 
$\tilde{r} = 1$; hence the $\tilde{r}$ axis is effectively divided into two regions where wave functions are localized 
below a threshold $\tilde{r}_{c}$ and extended above it.
 
The $D = 3$ bulk IPR curves in panel (c) of Fig.~\ref{fig:Fig12} differ in  
significant ways from the corresponding participation ratios calculated either for 1D or 2D systems.
In general, looking from $\tilde{r} = 0$ to $\tilde{r} = 1$, the bulk IPR initially
makes a sharp descent until intersecting the horizontal axis.  
The slope downward decreases with the hopping range $l$, and the normalized channel number  
where the bulk IPR becomes zero retreats to smaller $\tilde{r}$ with increasing scale $l$.  
For $\gamma \leq 1$, there is only a single mobility boundary; the IPR falls to zero, coincides with the abscissa, 
and does not climb again to finite values.  

On the other hand, although for larger $\gamma$ 
the extrapolated participation ratio curves initially descend as in $\gamma \leq 1$ and eventually 
become zero, in the upper $\tilde{r}$ range, the IPR begins to increase and is again 
finite.  Thus, for $\gamma > 1$, two mobility boundaries separate
 an intermediate range of energies where the participation ratio 
vanishes (corresponding to extended states) from energy regimes where wave functions are localized.
The width of the interval where the $\mathrm{IPR}_{0} = 0$   
constricts width increasing $\gamma$, though even for quite large values of the 
decay constant (i.e. for $\gamma = 5.5$), the extrapolated IPR seems at least to briefly touch the 
horizontal axis before rising again to finite values with a locally ``v'' shaped profile.  
Participation ratio curves corresponding to smaller $\gamma$ values have broader minima, but 
also have local symmetry about the same point in the vicinity of $\tilde{r} = 0.58$.

\begin{figure}
\includegraphics[width=.39\textwidth]{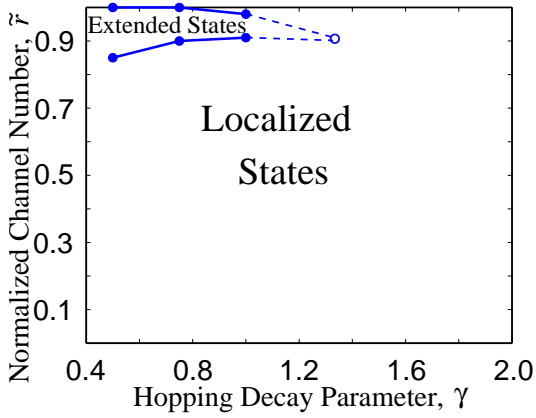}
\caption{\label{fig:Fig15} (Color Online) Phase portrait for 2D systems with the 
hopping decay parameter $\gamma$ on the abscissa and the channel number $\tilde{r}$ on  
the vertical axis.  The broken lines are extrapolated parts of the phase boundary.}
\end{figure}

\begin{figure}
\includegraphics[width=.39\textwidth]{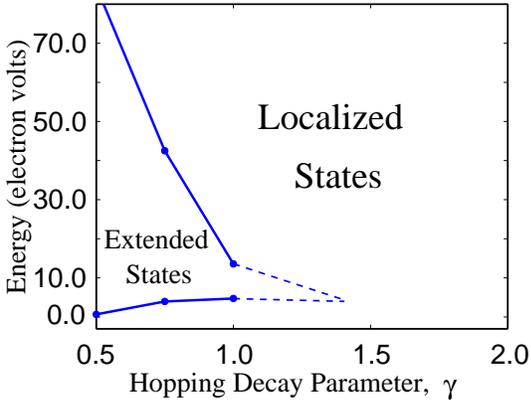}
\caption{\label{fig:Fig16} (Color Online) Phase diagram for the $D = 2$ case 
with the hopping decay parameter $\gamma$ on the abscissa and the eigenstate energy
(in electron volts) on the vertical axis.  The broken lines indicates extrapolated
parts of the phase boundary.}
\end{figure}

\begin{figure}
\includegraphics[width=.49\textwidth]{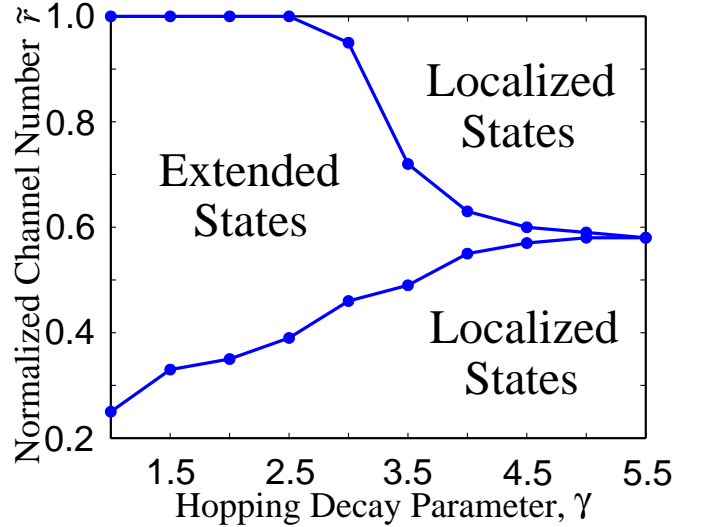}
\caption{\label{fig:Fig17} (Color Online) Phase portrait for 3D systems with the 
hopping decay parameter $\gamma$ on the abscissa and the channel number $\tilde{r}$ on the 
vertical axis.} 
\end{figure}

\begin{figure}
\includegraphics[width=.49\textwidth]{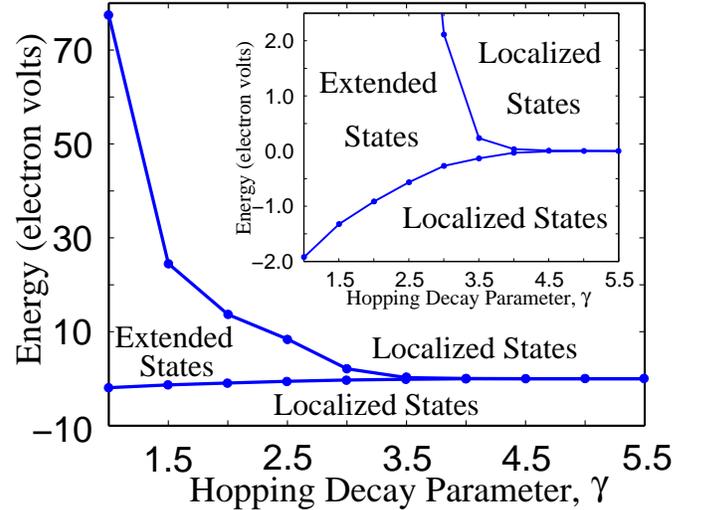}
\caption{\label{fig:Fig18} (Color Online) Phase diagram for the $D = 3$ 
case with the hopping decay parameter $\gamma$ on the abscissa and the 
eigenstate energy (in electron volts) on the vertical axis.  The graph inset is a closer 
view of the bifurcation region.}
\end{figure}

We use bulk participation ratios
to construct phase diagrams for the tight binding 
wave functions; for the 1D case, there is only a single phase with
the electronic states are localized in all cases, while phase portraits for $D = 2$ (shown in 
Fig.~\ref{fig:Fig15} and Fig.~\ref{fig:Fig16}) and $D = 3$ (displayed in Fig.~\ref{fig:Fig17} and 
Fig.~\ref{fig:Fig18}) show areas of extended states and phases in which all of the wave 
functions are localized.  For the 2D case, the wedge of extended states, broadest for 
$\gamma = 0.5$, tapers with increasing $\gamma$ and vanishes altogether in the 
vicinity of $\gamma = 1.5$, as indicated by the broken lines which are extrapolations.

For $D = 3$, the swath of extended states is quite broad for $\gamma \sim 1.0$, encompassing most of the 
eigenstates and the associated range of eigenstates energies.  As in the 2D case, a smaller portion 
of the wave functions are extended with increasing $\gamma$.  for moderate to large hopping integral     
decay lengths (i.e. for $\gamma \geq 3.0$), the band of extended states rapidly constricts with  
the extended phases asymptotically symmetric about $\tilde{r} = 0.58$.  Fig.~\ref{fig:Fig18} is the 
phase diagram with the vertical axis rendered in terms of energies instead of normalized rank $\tilde{r}$.  
Again, the interval of extended states becomes sharply narrower with increasing $\gamma$.  

\begin{figure}
\includegraphics[width=.49\textwidth]{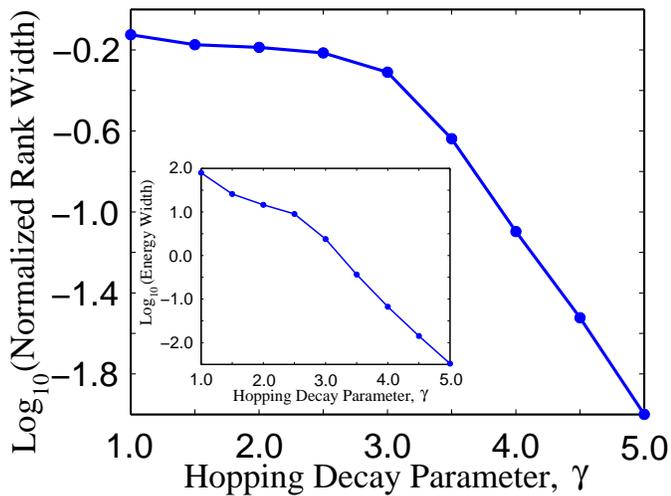}
\caption{\label{fig:Fig19} (Color Online) Logarithm of the 
width of the extended state region plotted versus the decay parameter $\gamma$.
The main graph shows the normalized rank width $w(\tilde{r})$, and the inset 
plot displays the width of the corresponding energy interval $w(E)$.}
\end{figure}

Although the extended state phase persists for large values of $\gamma$ (i.e. even for $\gamma > 5.0$),
the decrease of the width is very rapid, and we examine the possibility that the decrease may be exponential
in the hopping integral decay rate $\gamma$.
Fig.~\ref{fig:Fig19} displays a graph of base ten logarithm of the width of the phase where  
eigenstates are extended; the main plot shows $\log_{10}[w(\tilde{r})]$ with respect to $\gamma$, while the inset is a plot of 
$\log_{10}[w(E)]$ versus $\gamma$.  The asymptotically linear dependence 
of the logarithm of $w$, which is seen whether one considers the 
the width $w(E)$ of the energy interval or $w(\tilde{r})$  of the normalized rank,
is consistent with an exponential dependence  $w \propto  e^{-A \gamma} $ for moderate to large $\gamma$.

\section{conclusions and future work}

We have calculated the energy Density of States and, using the Inverse Participation 
Ratio,
we have examined the characteristics of electronic states in amorphous systems in one, 
two, and three dimensions for hopping matrix elements which decay exponentially in the separation  
distance between neighboring sites in the context of a tight binding model.  
We have calculated global IPR statistical distributions, which 
have a rich multimodal structure for systems in two and three dimensions
in contrast to the simple sharply peaked profiles consistently seen for $D = 1$.

Partitioning wave functions according to the normalized
eigenvalue rank $\tilde{r}$ has yielded channel averaged participation 
ratios, which we have shown to be representative of 
wave functions for a specific energy $E$ or normalized 
rank $\tilde{r}$.
By applying finite size scaling to each of the channel averages, we have obtained
participation ratios in the thermodynamic limit; using 
the latter, we have constructed phase portraits of the 
eigenstates with respect to localization.

In 1D, wave functions are strongly localized in all 
cases, whereas results for 2D indicate the presence of a 
critical decay parameter $\gamma_{c}$, with localization for 
$\gamma > \gamma_{c}$ and existence of extended states below $\gamma_{c}$.
In the 3D case, extended states also are admitted, and the 
wave functions with extended character occur even for 
quite large $\gamma$ values, although the interval of energies supporting 
extended states diminishes with $\gamma$, asymptotically scaling as $e^{-A \gamma}$ with increasing $\gamma$.  
The swath of extended states is flanked by regions where wave functions are localized, with the 
two interfaces interpreted as mobility boundaries.

In future studies, disorder schemes will be considered in which the severity of the disorder 
is tuned from mild to quite strong by perturbing a regular periodic crystal lattice and introducing
random perturbations $\delta$ in the positions of the hopping sites.  The disordering shifts $\delta$ 
may be introduced, e.g., from a Gaussian distribution with a RMS magnitude $\sigma$.  
Among the salient germane questions to be investigated in this manner is whether 
there is a threshold in typical displacement magnitudes where extended states may 
survive in $D=1$ and $D=2$ if random displacements in atomic positions are 
sufficiently small in relation to the crystal lattice constant.  Given the fragility of 
extended character in $D = 1$, one might predict that even a small random perturbation 
in the site positions from a periodic configuration might induce localization in a 
one dimensional lattice.  On the other hand, in two dimensions it may be that 
there is a perturbation level beyond which the disordering influence causes 
most or all of the states to be localized, with predominantly extended character 
below the perturbation threshold.

In the present study, we have concentrated on short-range couplings, as might be 
appropriate in an exchange type coupling scheme.  Nonetheless, it would be useful also 
to examine a power law decay to see if the severity of localization effects are reduced 
in one dimension, and if \textit{bona fide} extended states exist under these 
conditions.

\begin{acknowledgments}
Useful discussions with Euyheon Hwang, John Biddle, Bin Wang, and Sankar Das Sarma
are gratefully acknowledged.
The numerical analysis has been facilitated by use of the 360 node, 
2700 CPU University of Maryland, College Park HPCC parallel 
computing cluster.
\end{acknowledgments}


\end{document}